\documentclass[aip,jap,reprint,groupedaddress,showpacs]{revtex4-1}
\usepackage{graphicx}
\usepackage{latexsym,amssymb,amsmath}
\usepackage{bm}
\usepackage[caption=false]{subfig}
\usepackage{xcolor}

\usepackage[colorlinks=true,
urlcolor=blue,
linkcolor=blue,
citecolor=blue
]{hyperref}
\begin{document}

\title{Analysis of dependent scattering mechanism in hard-sphere Yukawa random media}
\author{B. X. Wang}
\affiliation{Institute of Engineering Thermophysics, Shanghai Jiao Tong University, Shanghai, 200240, People's Republic of China}


\author{C. Y. Zhao}
\email{Changying.zhao@sjtu.edu.cn}
\affiliation{Institute of Engineering Thermophysics, Shanghai Jiao Tong University, Shanghai, 200240, People's Republic of China}
\date{\today}

\begin{abstract}
The structural correlations in the microscopic structures of random media can induce the dependent scattering mechanism and thus influence the optical scattering properties. Based on our recent theory on the dependent scattering mechanism in random media composed of discrete dual-dipolar scatterers (\textit{Physical Review A 97, 023836}), in this paper, we study the hard-sphere Yukawa (HSY) random media, in order to further elucidate the role of structural correlations in the dependent scattering mechanism and hence optical scattering properties. Here, we consider charged colloidal suspensions, whose effective pair interaction between colloids is described by a screened Coulomb (Yukawa) potential. By means of adding salt ions, the pair interaction between the charged particles can be flexibly tailored and therefore the structural correlations are modified. It is shown that this strategy can affect the optical properties significantly. For colloidal $\mathrm{TiO_2}$ suspensions, the modification of electric and magnetic dipole excitations induced by the structural correlations can substantially influence the optical scattering properties, in addition to the far-field interference effect described by the structure factor. However, this modification is only slightly altered by different salt concentrations and is mainly because of the packing-density-dependent screening effect. On the other hand, for low refractive index colloidal polystyrene suspensions, the dependent scattering mechanism mainly involves the far-field interference effect, and the effective exciting field amplitude for electric dipole almost remains unchanged under different structural correlations. The present study has profound implications for understanding the role of structural correlations in dependent scattering mechanism.
\end{abstract}


\maketitle
\section{Introduction}
The propagation of light in disordered media is strongly affected by the micro/nanoscopic structures of the material, and the complicated interference of electromagnetic waves in such media may lead to intriguing transport phenomena, like coherent backscattering cone and Anderson localization \cite{wiersma1997localization,Segev2013}, position-dependent diffusion constant \cite{kopPRL1997}, sub-diffusive \cite{sebbahPRB1993} and super-diffusive \cite{barthelemyNature2008,bertolottiPRL2010} transport behaviors of light, etc. Moreover, disordered media with engineered micro/nano-structures exhibit promising  applications such as quantum optics \cite{garciaAnnPhys2017}, solar energy harvesting and conversion \cite{Vynck2012,Fang2015JQSRT,Liew2016ACSPh,liuJOSAB2018}, random lasers \cite{Cao1999,Wiersma2008,linACSPhoton2017}, radiative cooling \cite{zhaiScience2017,baoSEMSC2017} and structural coloration \cite{xiaoSciAdv2017}, etc. Particularly, disordered media composed of discrete scatterers, such as materials made of randomly packed zinc oxide ($\mathrm{ZnO}$), silica ($\mathrm{SiO_2}$) and titanium dioxide ($\mathrm{TiO_2}$) nanoparticles, are widely studied and used for manipulation light propagation due to the easiness of fabrication and the flexibility of controlling \cite{wiersma1997localization,Storzer2006,zhaiScience2017,baoSEMSC2017,linACSPhoton2017,xiaoSciAdv2017}. For such media, the radiative transport properties, including the scattering coefficient $\kappa_s$ (or the corresponding scattering mean free path $l_s=1/\kappa_s$) and the asymmetry factor $g$, are usually predicted by using the independent scattering approximation (ISA) under the framework of the phenomenological radiative transfer equation (RTE) \cite{VanRossum1998,mishchenko2006multiple,tsang2004scattering,sheng2006introduction}, which is valid only when the scatterers are far-apart from each other and no long-range positional correlations exist. 

When the concentration of particles rises, where inter-particle distance is comparable with or even smaller than the incident wavelength, the scattered waves from different scatterers begin to interfere with each other, leading to the failure of ISA \cite{garciaPRA2008,wangIJHMT2015,Naraghi2015}. There are continuous investigations of the so-called dependent scattering mechanism in the last a few decades \cite{tien1987thermal,durantJOSAA2007,garciaPRA2008,nguyenOE2013,wangIJHMT2015,Naraghi2015,maJQSRT2017,wangPRA2018}, in order to correctly predict the interference effects. One important factor that leads to the dependent scattering mechanism is known as the structural correlations, which describe possible reminiscence of order (usually short- or medium-ranged) existing in the spatial variation of the dielectric constant in disordered media \cite{tsang2004scattering2}. They can lead to definite phase differences among the scattered waves \cite{laxRMP1951,laxPR1952,fradenPRL1990,tsang2004scattering2,rojasochoaPRL2004,Froufe-PerezPNAS2017,liuJOSAB2018}, and hence result in remarkable impacts on the transport and scattering properties of light. The most well-known type of structural correlations is the hard-sphere positional correlation in disordered media consisting of pure hard spheres without any additional inter-particle interactions \cite{wertheimPRL1963,fradenPRL1990}. This is due to the fact that hard spheres can not deform or penetrate into each other, and the structural correlations emerge when the spheres are densely packed (usually under volume fraction $f_v>5\%$). Nevertheless, for other kinds of specially-designed structural correlations, even if the concentration of particles is not very high, the strong positional correlation will give rise to very significant interference phenomena, for instance, in the so-called short-ranged ordered hyperuniform media\cite{leseurOptica2016,Froufe-Perez2016,Froufe-PerezPNAS2017}.

Here we deal with the structural correlations of the hard-sphere Yukawa (HSY) type, which are usually encountered in highly charged colloidal suspensions as well complex (dusty) plasmas \cite{kalmanPRE2013}. For highly like-charged colloidal suspensions, the pair potential between colloids is not the pure Coulomb-repulsive-type but becomes a Yukawa-type potential due to Debye screening effect \cite{robbinsJCP1988,rojasochoaPRL2004,tataSSC2006,bresselJSQRT2013,kalmanPRE2013}. The screening strength can be easily tailored by adding extra salt ions \cite{robbinsJCP1988,tataSSC2006,kalmanPRE2013}. Therefore, the structural correlations induced by the pair potential in such colloids can be modified accordingly. This feature of charged colloidal suspensions provides us a realistic model system to investigate the role of structural correlations in the dependent scattering mechanism. Actually, manipulating the optical scattering properties of such colloidal suspensions by tuning the electrostatic interaction therein attracts an everlasting interests in the fields of light scattering \cite{bresselJSQRT2013}, photonics \cite{kolleADMA2017}, mesoscopic physics \cite{rojasochoaPRL2004} and polymer science \cite{huangLangmuir2002}, etc. It is also a promising way to realize strong light localization  \cite{rojasochoaPRL2004}, photonic bandgaps \cite{garciaADMA2007}, structural coloration \cite{HanADMA2012}, radiative cooling \cite{atiganyanunACSPhoton2018} and other potential applications if the electrostatic potential is well-controlled to realize desirable structural correlations.  Nevertheless, the dependent scattering mechanism induced by such specific structural correlations and how it influences optical properties of such media have not been extensively studied yet \cite{bresselJSQRT2013}. 

In this paper, based on our recently developed rigorous theory  of dependent scattering mechanism in random media consisting of dual-dipolar particles \cite{wangPRA2018}, we study how the structural correlations induce and affect the dependent scattering mechanism in hard-sphere Yukawa random media, namely, colloidal suspensions. Apart from the conventional far-field interference effect described by the static structure factor, we also show that the modification of electric and magnetic dipole excitations due to structural correlations is also important. By comparing two different kinds of colloidal particles, $\mathrm{TiO_2}$ and polystyrene (PS), we demonstrate that the refractive index contrast is a crucial factor that interplays with the structural correlations. Our study is promising in understanding and manipulating the optical scattering properties of dual-dipolar random media. 

\section{Model}\label{model}
In this paper, we will consider aqueous colloidal suspensions consisting of highly like-charged identical dual-dipolar spherical particles. The term ``dual-dipolar" here means that the optical response of an individual particle only includes electric and magnetic dipoles \cite{schmidtPRL2015}. We will only work on the so-called ``fluid" phase of such suspensions and do not take the glassy or solid (crystalline) phases due to structural ordering into account by appropriately choosing the system parameters \cite{robbinsJCP1988,meijerJCP1991,crockerPRL1994,weissmanScience1996,tataSSC2006,rodriguezPCL2016}. In this circumstance, the colloidal suspensions are random media, which only exhibit short-range order and has no long-range order. We also assume all the particles are isotropic, homogeneous and hard spheres with a radius of $a$. Their positions can be regarded as fixed since they are stabilized by the charges and move much slower than the electromagnetic waves \cite{denkovPhysicaA1992,mishchenko2006multiple,mishchenko2014electromagnetic}. Such random media are also supposed to be statistically homogeneous and isotropic. 
\begin{figure}[htbp]
	\centering
	\includegraphics[width=0.6\linewidth]{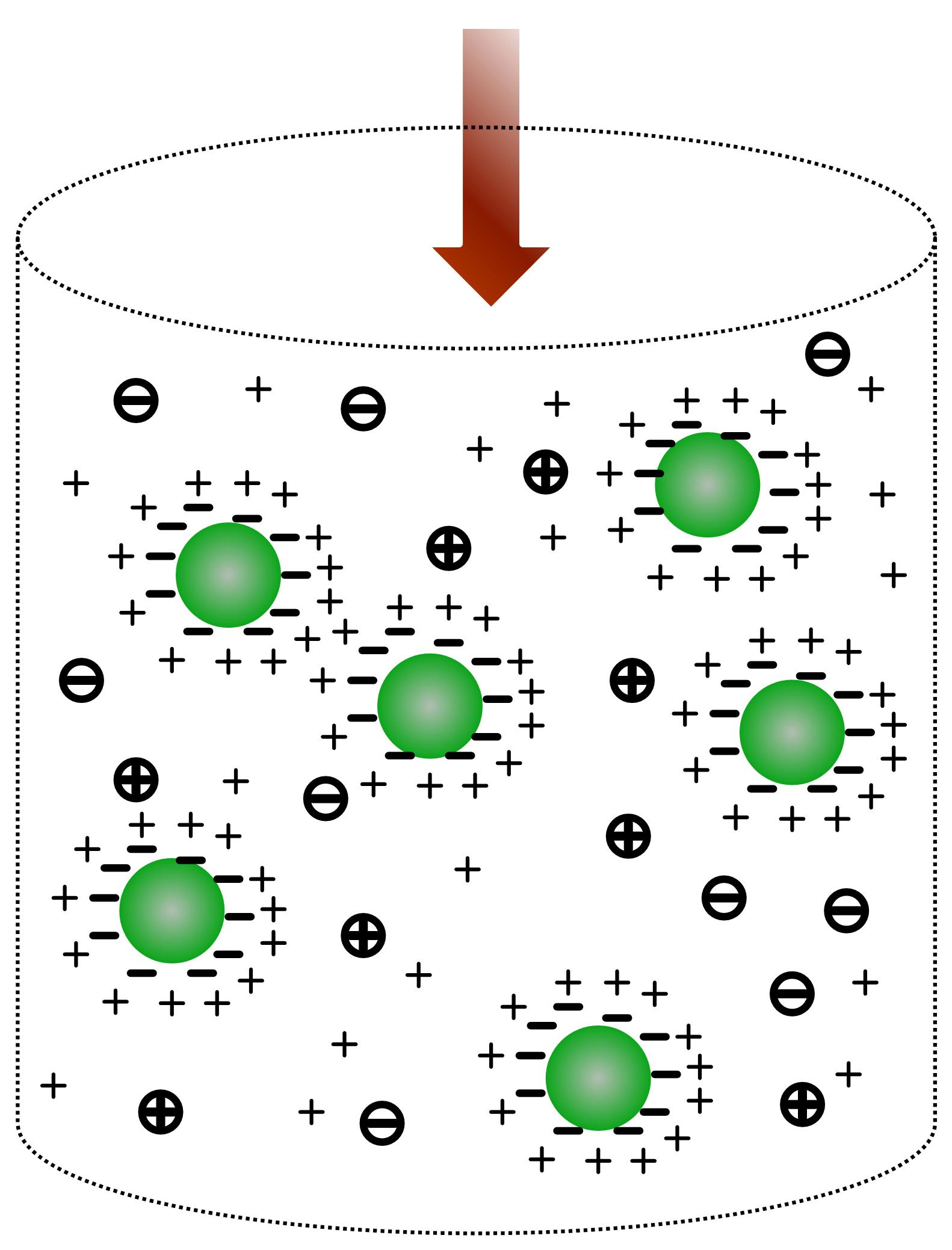}
	\caption{A schematic of the colloidal suspensions studied in this paper. The dotted lines stand for an imaginary boundary of the suspensions. This system constitutes negatively charged colloidal particles (large spheres), along with small counterions (+) and additional salt ions ($\oplus$ and $\ominus$). The propagation direction of incident light is denoted by the thick arrow.}
	\label{system_config}
\end{figure}
\subsection{Hard-sphere Yukawa system and pair correlation function}
Here, the aqueous colloidal suspensions consist of highly negatively charged monodisperse colloidal particles (or called macroions) in a solvent and corresponding tiny counterions to make the suspension electrically neutral. The counterions can screen the pure Coulomb potential, and there could be additional salt ions in the solvent to give rise to further screening effects to control the inter-particle potential. The schematic of such suspensions is shown in Fig.\ref{system_config}. According to the Derjaguin-Landau-Verwey-Overbeek (DLVO) theory, the effective pair interaction between charged particles in such suspensions is of the repulsive screened Coulomb form (or HSY form). This is because the electrostatic potential is generally much larger than thermal energies at room temperature (measured by $k_BT$ where $k_B$ is the Boltzmann constant and $T$ is the temperature of the system), and thus the weakly attractive van der Waals force plays little role in the time-averaged structural correlations \cite{hayterMP1981,rowlinsonPA1989}. This HSY potential reads \cite{huangLangmuir2002,messinaPRL2003}, 
\begin{equation}\label{HSY_eq} 
U_{\mathrm{HSY}}(\mathbf{r})=\begin{cases}
\infty &{\text{when } 0<r<d,}\\
U_0\frac{\exp[-\kappa_D(r-d)]}{r/d} &{\text{when } r\geq d,}
\end{cases}
\end{equation}
where $d=2a$ is the diameter of the particles. $U_0$ is the contact potential between a pair of particles:
\begin{equation}
U_0=\frac{Z_\mathrm{eff}^2e^2}{4\pi\varepsilon_r\varepsilon_0d(1+\kappa_Da)^2},
\end{equation}
where $Z_\mathrm{eff}$ is the effective charge of the charged particles (macroions), $e$ is the elementary charge, $\varepsilon_0$ is the electric permittivity in vacuum. For water at room temperature the static permittivity is $\varepsilon_r\approx80$. Here $\kappa_D$ is the inverse Debye-H\"uckel screening length, which is related to the concentration of counterions and additional salt ions \cite{messinaPRL2003,rojasochoaPRL2008}:
\begin{equation}
\kappa_D=\sqrt{\frac{e^2(Z_\mathrm{eff}n_p+n_s)}{\varepsilon\varepsilon_rk_BT}},
\end{equation}
where $n_p$ and $n_s$ are the number densities of particles and additional salt ions (here we account for monovalent ions), respectively. Therefore, the screening length can be flexibly tuned by the concentration of additional salt ions. Direct observation of Eq.(\ref{HSY_eq}) tells us that for very small $\kappa_D$, the HSY potential approaches the Coulomb potential, while for very large $\kappa_D$, it becomes a hard-sphere potential. 

Here we consider two types of charge-stabilized colloidal suspensions in room temperature which can be easily prepared in experiments. The first is charged polystyrene (PS) nanoparticle suspensions \cite{huangLangmuir2002,bresselJSQRT2013}, and the other is a colloidal suspension composed of charged rutile ($\mathrm{TiO_2}$) nanoparticles (NPs) \cite{xiaCAAJ2017,duran-ledezmaAO2018}. In both cases, we assume the solvent is water. The radius of a spherical nanoparticle is $a=100\mathrm{nm}$. The wavelength of incident light is set to be $\lambda=633\mathrm{nm}$, where the refractive index of $\mathrm{TiO_2}$ is $m_\mathrm{TiO_2}=2.58$ and that of PS is $m_\mathrm{PS}=1.59$. The effective charge of each particle is assumed to be fixed at $Z_\mathrm{eff}=100$. Note the effective charge corresponds to a charge renormalization procedure in applying the DLVO theory, which is related to the complicated electro-kinetic interaction among counterions, salt ions and the charged surfaces of the particles. Thus in reality it changes with the concentration of additional salt ions \cite{soodSSP1991,rojasochoaPRL2008}, which can be predicted by several models including the Poisson-Boltzmann-cell (PBC) model
and the Poisson-Boltzmann-jellium (PBJ) approximation \cite{soodSSP1991}. However, this is out the scope of the present paper. Since we are only concerned with the dependent scattering mechanism in the optical aspect, the fixed effective charge in essence does not affect the understanding on the multiple scattering physics. In fact, we can tune the bare charge of the macroions accordingly to keep the effective charge $Z_\mathrm{eff}=100$ when the concentration of small counterions changes in order to realize the situations in our theoretical calculation.

On the basis of HSY potential, solving the Ornstein-Zernike (OZ) equation using the mean spherical approximation (MSA) Yukawa system gives the static structure factor of such system \cite{blumJSP1980,ginozaMP1990,bresselJSQRT2013}:
\begin{equation}
S(\mathbf{q})=\left\{[1-6f_vZ_1(u)]^2+36f_v^2Z_2^2(u)\right\}^{-1},
\end{equation}
where $\mathbf{q}$ is the reciprocal vector, $f_v$ is the volume fraction of particles and $u=2qa$. For isotropic media, the structure factor only depends on $q=|\mathbf{q}|$. $Z_1(u)$ and $Z_2(u)$ can be calculated analytically from the formulas listed in Appendix \ref{MSA}. In the meanwhile, from the structure factor, we are able to calculate the pair correlation function (PCF) as
\begin{equation}\label{ftransform}
h_2(\mathbf{r})=h_2(r)=\int_{-\infty}^{\infty}H(\mathbf{q})\exp{(i\mathbf{q}\cdot\mathbf{r})}d\mathbf{r},
\end{equation}
where $H(\mathbf{q})=[S^{-1}(\mathbf{q})-1]/[n_0(2\pi)^3]$ is the pair correlation function in momentum space. The PCF and corresponding structure factor describe two-particle statistics in the structural correlations of a random medium, providing a theoretical basis to explore the effect of structural correlations on the dependent scattering mechanism and thus the optical scattering properties. 

\begin{figure}[htbp]
	\centering
	\subfloat{	
		\label{fv01Sq}
		\includegraphics[width=0.8\linewidth]{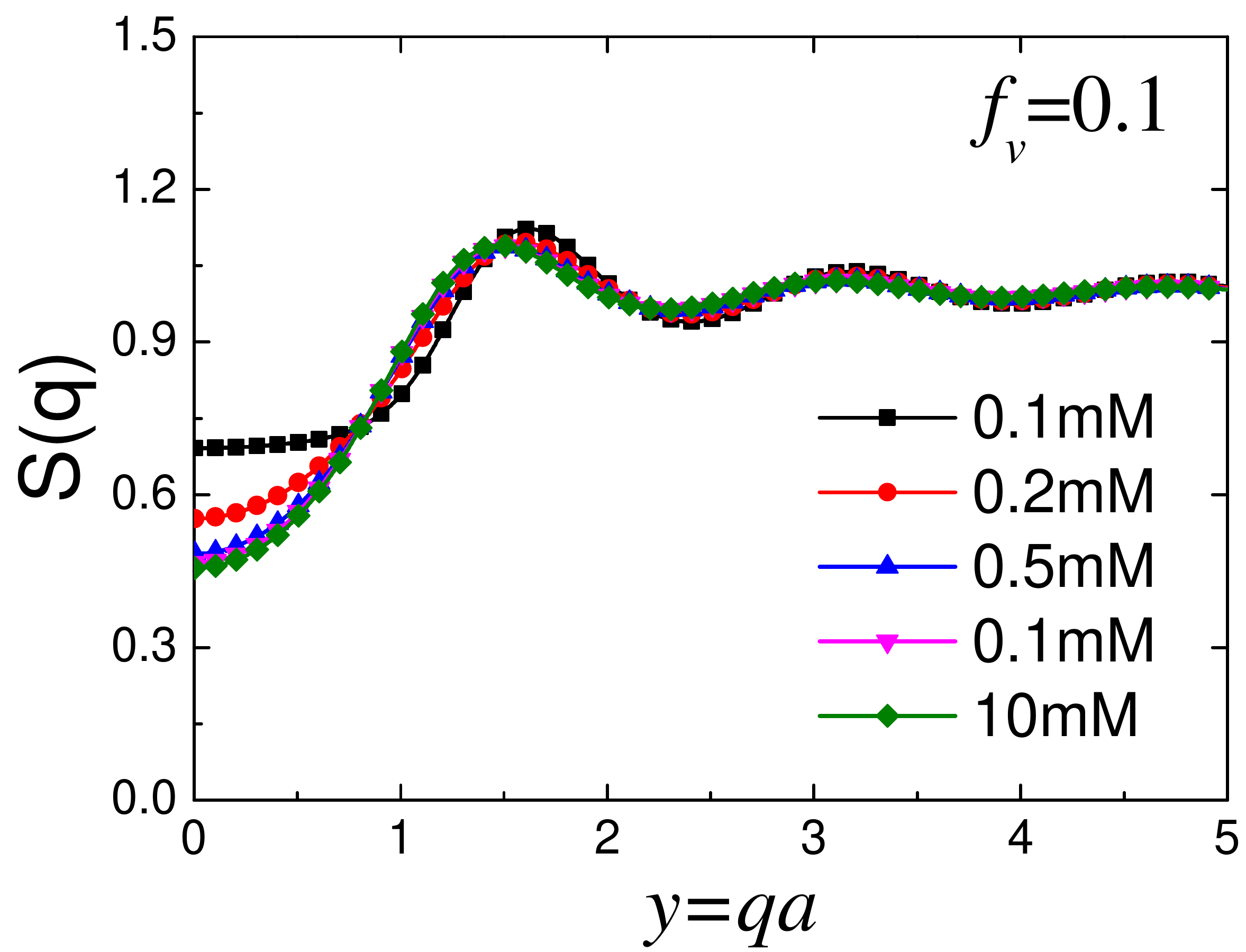}
		
	}
	\hspace{0.01in}
	\subfloat{	
		\label{fv02Sq}
		\includegraphics[width=0.8\linewidth]{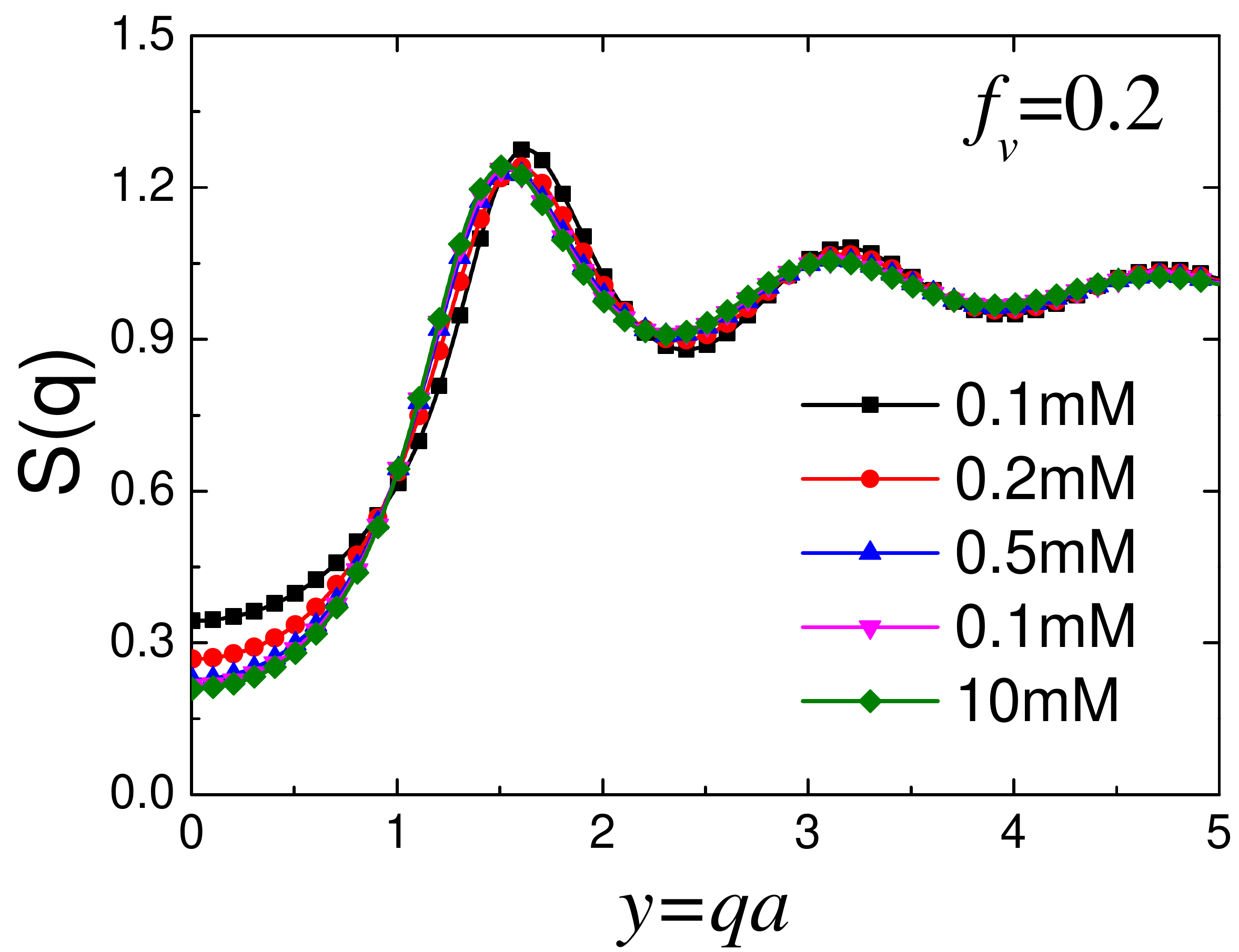}
	}
	\caption{The structure factor $S(q)$ as a function of wavelength for HSY random media with different salt ion concentrations. (a) $f_v=0.1$; (b) $f_v=0.2$.}\label{Sq}
\end{figure}

Actually, the structure factor determines the far-field interference effect in dependent scattering mechanism. This fact is shown by many authors and constitutes the first order dependent-scattering correction to the ISA \cite{fradenPRL1990,mishchenkoJQSRT1994,huangLangmuir2002,rojasochoaPRL2004,conleyPRL2014,liuJOSAB2018}. It in principle describes the far-field interference between first-order scattered waves of different particles, and is also called the interference approximation (ITA) \cite{dickJOSAA1999} or the collective scattering correction \cite{Naraghi2015}. In Fig.\ref{Sq}, we show the structure factor $S(q)$ of different random systems as a function of $y=qa$ for particle volume fractions $f_v=0.1$ and $f_v=0.2$, for additional salt concentration $c_s=0.1,0.2,0.5,1,10 \mathrm{mM}$. The number density of salt ions is calculated through $n_s=2000N_Ac_s$ where $N_A$ is the Avogadro constant. This rise of salt concentration makes the normalized inverse screening length $\kappa_Dd$ increases monotonously from 6.6 to 65.1597, approaching the hard-sphere limit (i.e., infinitely short interacting potential). It can be seen from Fig.\ref{Sq} that the salt concentration mainly affects the forward scattering intensity (for low $q$), while for large $q$ ($y=qa\gtrsim1$), the difference is very slight. This is because we are working on a very small area in the phase diagram of Yukawa system, in order to guarantee the random media are in a ``fluid" state, where no glassy or solid (crystalline) states can emerge \cite{robbinsJCP1988}. The latter can significant modify the structure factor, which are out of the scope of the present paper. Although the difference in the structural correlations is only manifested in the low-$q$ zone of structure factor, we will show later this can still influence the optical scattering properties remarkably.

\subsection{Effective propagation constant and scattering phase function}\label{theory1}
By now knowing the PCFs under different structural correlations, we are able to calculate the optical scattering properties based on our recently developed theory considering the dependent scattering mechanism in dual-dipolar random media \cite{wangPRA2018}. This theory provides analytical expressions for the effective propagation constant, scattering coefficient and phase function for the random media, by exploiting the multipole expansion method and quasicrystalline approximation (QCA) for the Foldy-Lax equations (FLEs), which rigorously describe multiple scattering of electromagnetic waves for a system of discrete scatterers. In this theory, the effective propagation constant $K$ in the random media is solved from the following equation
\begin{equation}\label{dispersion_relate_eq}
[1-n_p\tilde{A}_{1111}(K)b_1][1-n_p\tilde{A}_{1111}(K)a_1]-n_p^2\tilde{A}_{1211}^2a_1b_1=0
\end{equation}
in the upper complex plane of $K$. $a_1$ and $b_1$ are Mie coefficients indicating electric and magnetic dipoles respectively. The elements of $\tilde{\mathbf{A}}(\mathbf{K})$ appearing in the above equations are obtained through an integral involving the PCF, and can be found in Ref.\cite{wangPRA2018}. Moreover, our theory predicts the effective exciting field amplitudes for electric ($C_{12}$) and magnetic dipoles ($C_{11}$) beyond the traditional ITA, to account for the modification of exciting field impinging on the particles due to dependent scattering mechanism \cite{wangPRA2018}.  $C_{12}$ and $C_{11}$ can be solved from the following equations:
\begin{equation}\label{extinction_theorem}
K^2-k^2=\frac{6\pi in_p}{k}(a_1C_{12}+b_1C_{11}).
\end{equation}
\begin{equation}\label{c_eq1}
\begin{split}
[1-n_p\tilde{A}_{1111}(K)b_1]C_{11}-n_p\tilde{A}_{1211}(K)a_1C_{12}=0,
\end{split}
\end{equation}
where $k=2\pi m_b/\lambda$ is the wave number in the background medium with a refractive index $m_b$. Here the background medium is water with $m_b=1.35$. The size parameter of the colloidal particle is $x=ka=1.89$, and in $\mathrm{TiO_2}$ particles both electric and magnetic dipoles are excited while in PS particles only electric dipole contributes to the electromagnetic response according to Mie theory. The present theoretical framework enables us to determine how the structural correlations play a role in the modification of the electric and magnetic dipole excitations. This effect of dependent scattering mechanism is seldom analyzed for colloidal suspensions and will be discussed in this paper in detail.

After calculating $K$, $C_{12}$ and $C_{11}$, further taking the on-shell and far-field approximations \cite{wangPRA2018}, we obtain the different scattering coefficient as \cite{wangPRA2018}
\begin{equation}\label{pf_qca2}
\begin{split}
&\frac{d\kappa_\text{s}}{d\theta_\text{s}}=\frac{9n_p}{4k^2}S(\mathbf{p}'-\mathbf{p})\\
&\times\Big[|a_1C_{21}\pi_n(\cos\theta_\text{s})+b_1C_{11}\tau_n(\cos\theta_\text{s})|^2\\
&+|b_1C_{11}\pi_n(\cos\theta_\text{s})+a_1C_{21}\tau_n(\cos\theta_\text{s})|^2\Big],
\end{split}
\end{equation}
where $\theta_\text{s}$ indicates the polar scattering angle between incident direction $\mathbf{p}$ and scattering direction $\mathbf{p}'$, and the dependency on azimuth angle is integrated out due to azimuth symmetry in random media. The functions $\tau_n(\cos\theta_\text{s})$ and $\pi_n(\cos\theta_\text{s})$ are derived from Legendre functions and can be found in Ref.\cite{wangPRA2018}. Note here different from conventional ITA, the argument in the structure factor is given by $|\mathbf{p}'-\mathbf{p}|=\sqrt{K^2+k^2-2Kk\cos\theta_s}$, which accounts for the propagation of effective mode in the random media. From Eq.(\ref{pf_qca2}) it can be observed that in the present theory, in addition to the contribution of structure factor, the structural correlations also modify the effective exciting field amplitudes $C_{12}$ and $C_{11}$, which are both assumed to be equal to unity in ISA and ITA. If they are different from unity, the effective propagation constant (i.e., the effective refractive index), scattering coefficient and phase function will be substantially affected. 

It should be noted that the recurrent scattering effect is not included in our theoretical model\cite{Aubry2014PRL}. This effect implies the contribution of those multiple scattering trajectories that visit the same particle more than once and form closed loops. It is significant for extremely strong scattering media, for instance, cold atomic clouds \cite{Cherroret2016}, and can occur where no structural correlations exist \cite{vanTiggelenPRE2006}. Since here we only deal with structural correlations in moderately scattering media, the recurrent scattering effect is assumed to be small, which is also beyond the scope of this paper and will not be discussed.

\section{Results and Discussion}\label{results}
\begin{figure}[htbp]
	\centering
	\subfloat{	
		\label{kappasTiO2Z200}
		\includegraphics[width=0.8\linewidth]{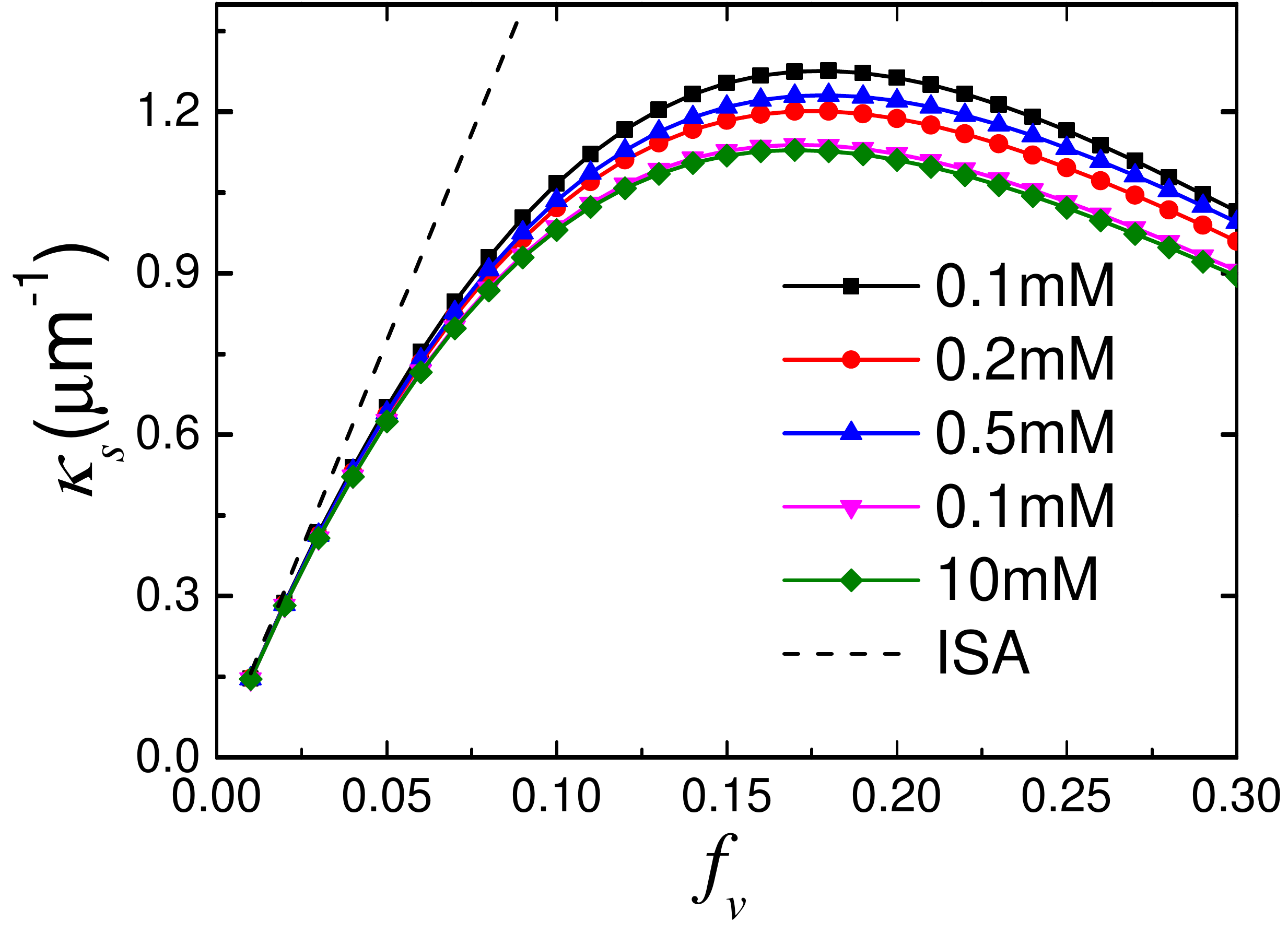}
		
	}
	\hspace{0.01in}
	\subfloat{	
		\label{ggTiO2Z200}
		\includegraphics[width=0.8\linewidth]{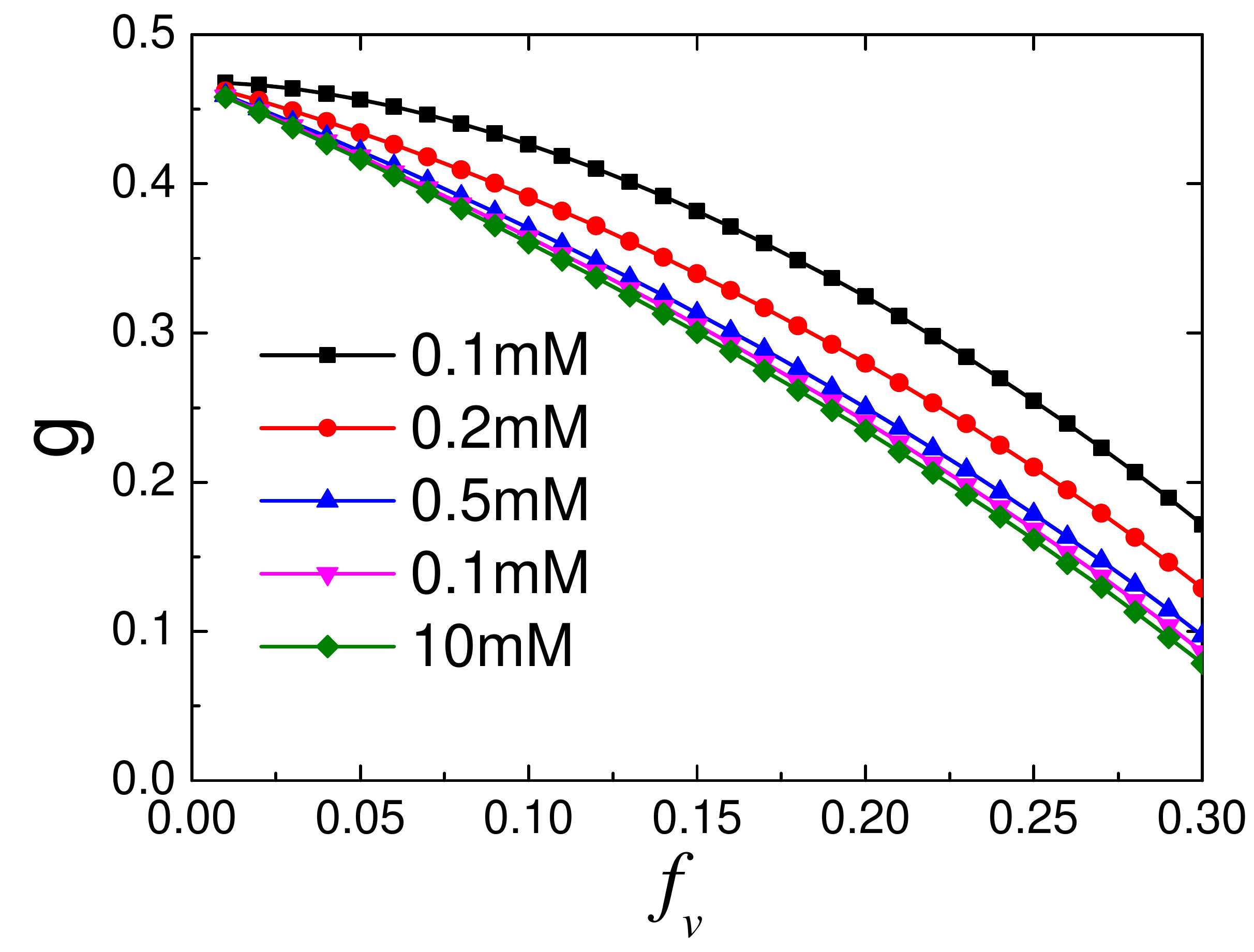}
	}
	\caption{ (a) The scattering coefficient $\kappa_s$ and (b) asymmetry factor $g$ as a function of particle volume fraction for $\mathrm{TiO_2}$ colloidal suspensions with different salt ion concentrations.}\label{TiO2Z200}
\end{figure}

\begin{figure}[htbp]
	\centering
	\subfloat{	
		\label{C12TiO2Z200}
		\includegraphics[width=0.8\linewidth]{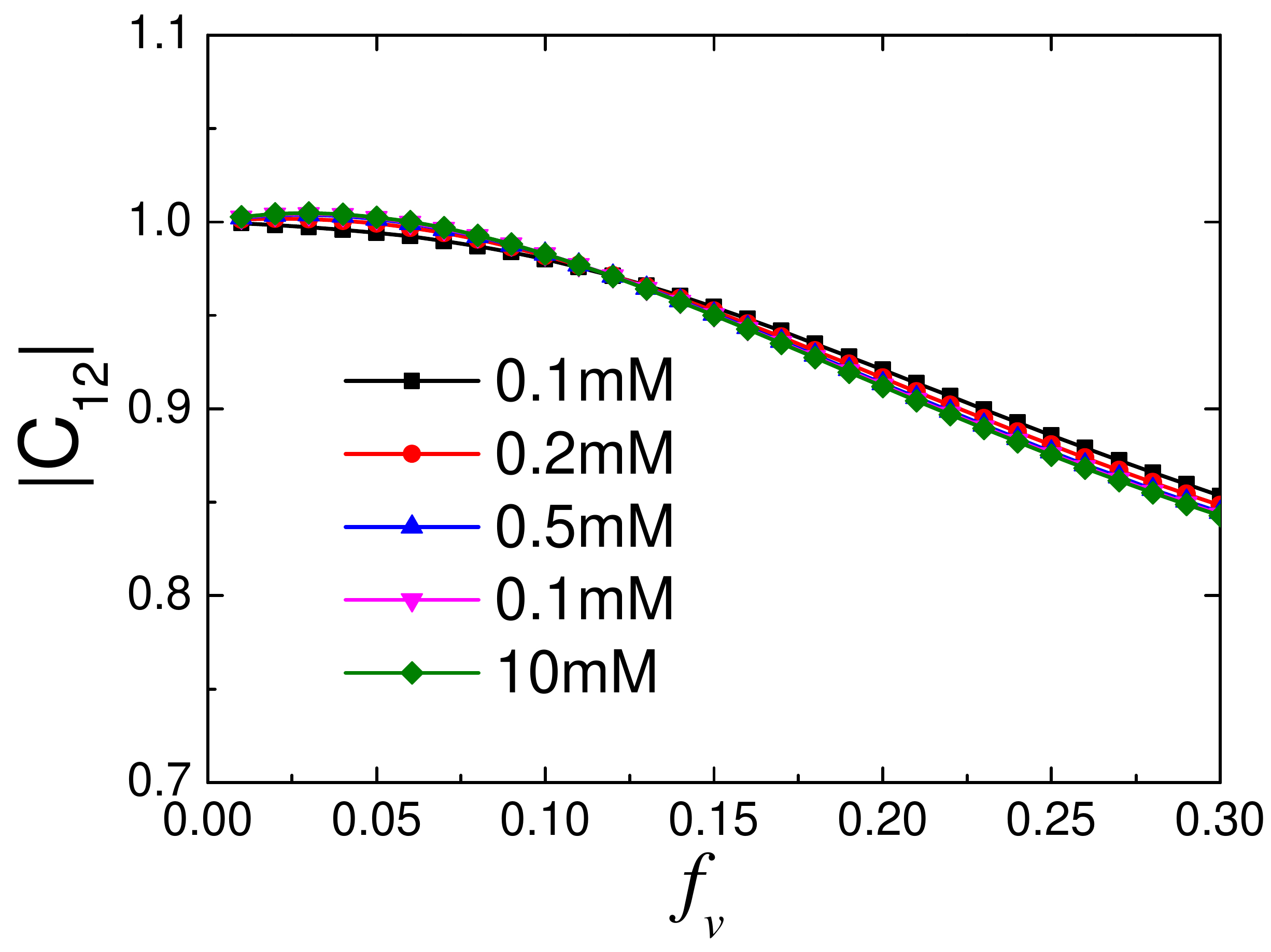}
		
	}
	\hspace{0.01in}
	\subfloat{	
		\label{C11TiO2Z200}
		\includegraphics[width=0.8\linewidth]{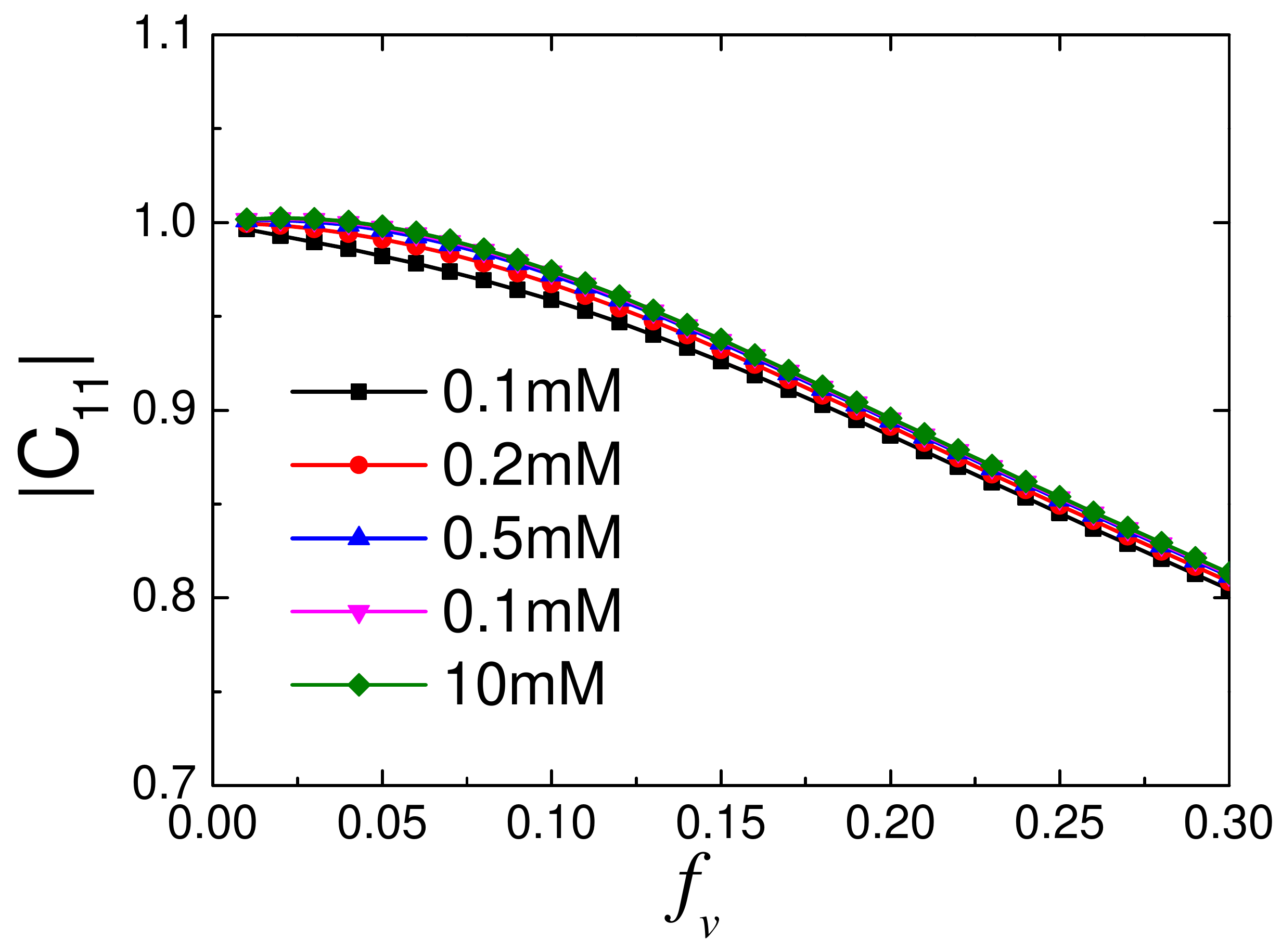}
	}
	\caption{The effective exciting field amplitudes (a) $|C_{12}|$ for electric dipole and (b) $|C_{11}|$ for magnetic dipole as a function of particle volume fraction for $\mathrm{TiO_2}$ colloidal suspensions with different salt ion concentrations.}\label{CTiO2Z200}
\end{figure}

As mentioned in the previous section, here we study $\mathrm{TiO_2}$ and PS colloidal suspensions, where effective surface charge of a particle is kept to be $Z_\mathrm{eff}=100$ and additional salt concentration is tuned as $c_s=0.1,0.2,0.5,1,10 \mathrm{mM}$. For the $\mathrm{TiO_2}$ colloidal suspensions, the scattering coefficient $\kappa_s$ and asymmetry factor $g$ as a function of the particle volume fraction are shown in Figs.\ref{kappasTiO2Z200} and \ref{ggTiO2Z200}, along with the effective field amplitudes shown in Fig.\ref{CTiO2Z200}. In Fig.\ref{kappasTiO2Z200}, it is observed that the scattering coefficient gradually increases with volume fraction for $f_v\lesssim0.15$. This increase is mainly due to the growth of particle density $n_p$, where the dependent scattering effect is not strong enough to suppress it. While when $f_v\gtrsim0.15$, the scattering coefficient lowers down due to the intensification of dependent scattering mechanism at larger particle densities. The scattering coefficient calculated by ISA is also shown in dashed line for comparison, which predicts a linear increase with volume fraction. The deviation of the result of ISA from that of dependent scattering model thus grows rapidly with particle density. (The result of ISA above $f_v=0.1$ is not shown for the compactness of the figure.) The dependent scattering effects here are manifested in both structure factor shown in Fig.\ref{Sq} and the effective field amplitudes shown in Fig.\ref{CTiO2Z200}. Regarding the structure factor, at volume fraction $f_v=0.2$ the forward scattering intensity (i.e., where $y=qa$ is small) is significantly reduced compared to that in $f_v=0.1$. Moreover, from Fig.\ref{CTiO2Z200}, we can recognize that at higher particle densities, both $|C_{12}|$ and $|C_{11}|$ monotonously decrease, from unity in the dilute limit to only 0.8 at $f_v=0.3$. This can be understood as a screening effect, that is that each particle witnesses a weaker exciting field amplitude than the incident field due to the screening of nearby particles. Screening effect is enhanced at higher particle densities. Traditionally this screening effect and resultant smaller refractive index contrast between particles and background is considered phenomenologically by using an effective index (usually by Maxwell-Garnett theory \cite{Liew2011}) in the denominator of Eq.(\ref{pf_qca2}), while in our model we take this into account more rigorously. In Fig.\ref{ggTiO2Z200}, it can also be understood that due to the weakening of forward scattering intensity shown in structure factor, the asymmetry factor decreases with the volume fraction grows.

When the salt concentration is increased, the scattering coefficient substantially decreases for $f_v>0.15$ where dependent scattering effects becomes remarkable. The asymmetry factor also reduces with higher salt concentration. However, the effective field amplitudes $|C_{12}|$ and $|C_{11}|$ keep almost unchanged under different salt concentrations. We can hence conclude that here the differences in optical scattering properties are mainly due to the far-field interference effect from different structural correlations, whereas the effective field amplitudes are almost unaffected. However, we should keep in mind this conclusion may be only valid for present random media which cover only a very small range in the entire phase diagram of HSY systems.

\begin{figure}[htbp]
	\centering
	\subfloat{	
		\label{kappasPSZ200}
		\includegraphics[width=0.8\linewidth]{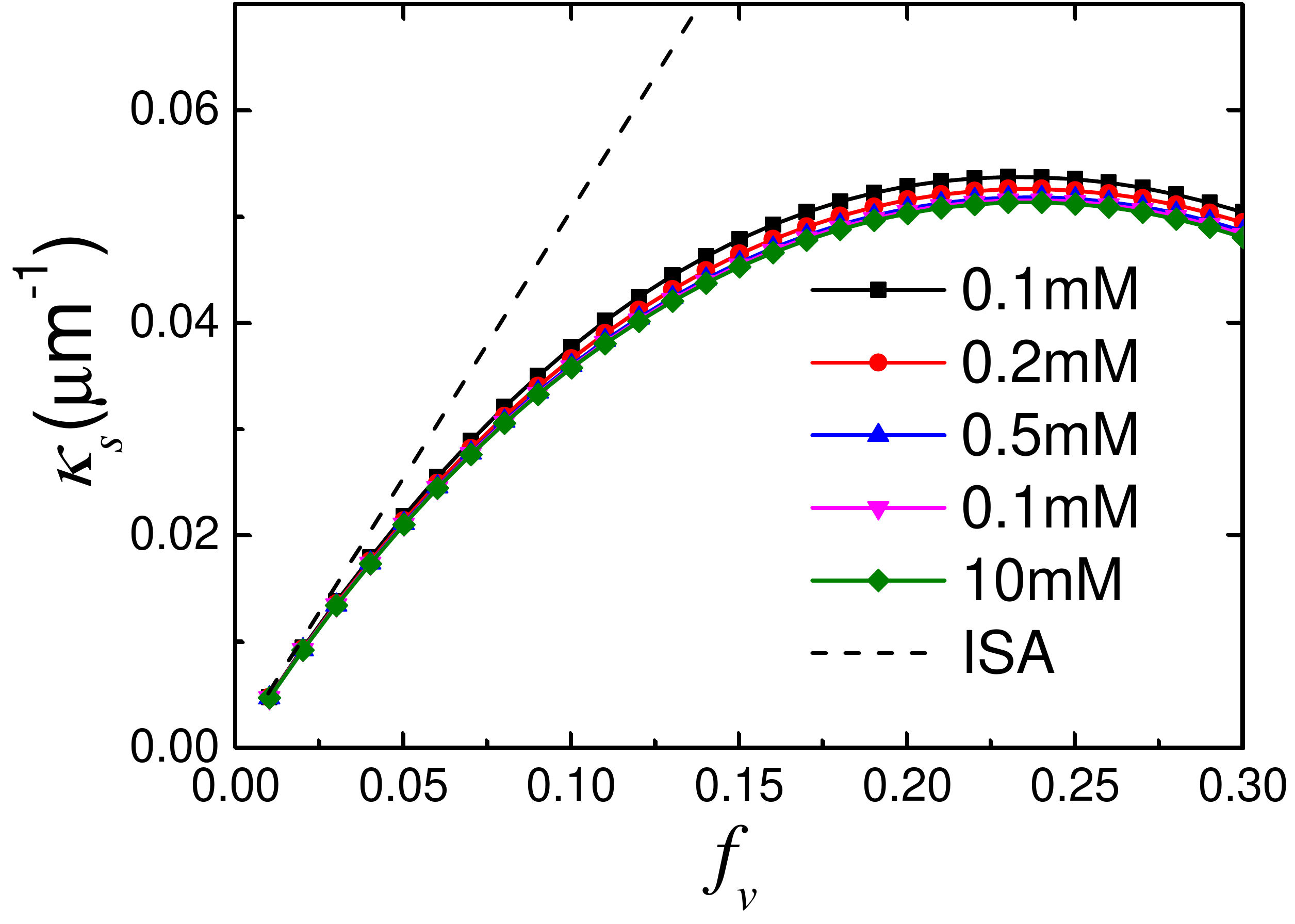}
		
	}
	\hspace{0.01in}
	\subfloat{	
		\label{ggPSZ200}
		\includegraphics[width=0.8\linewidth]{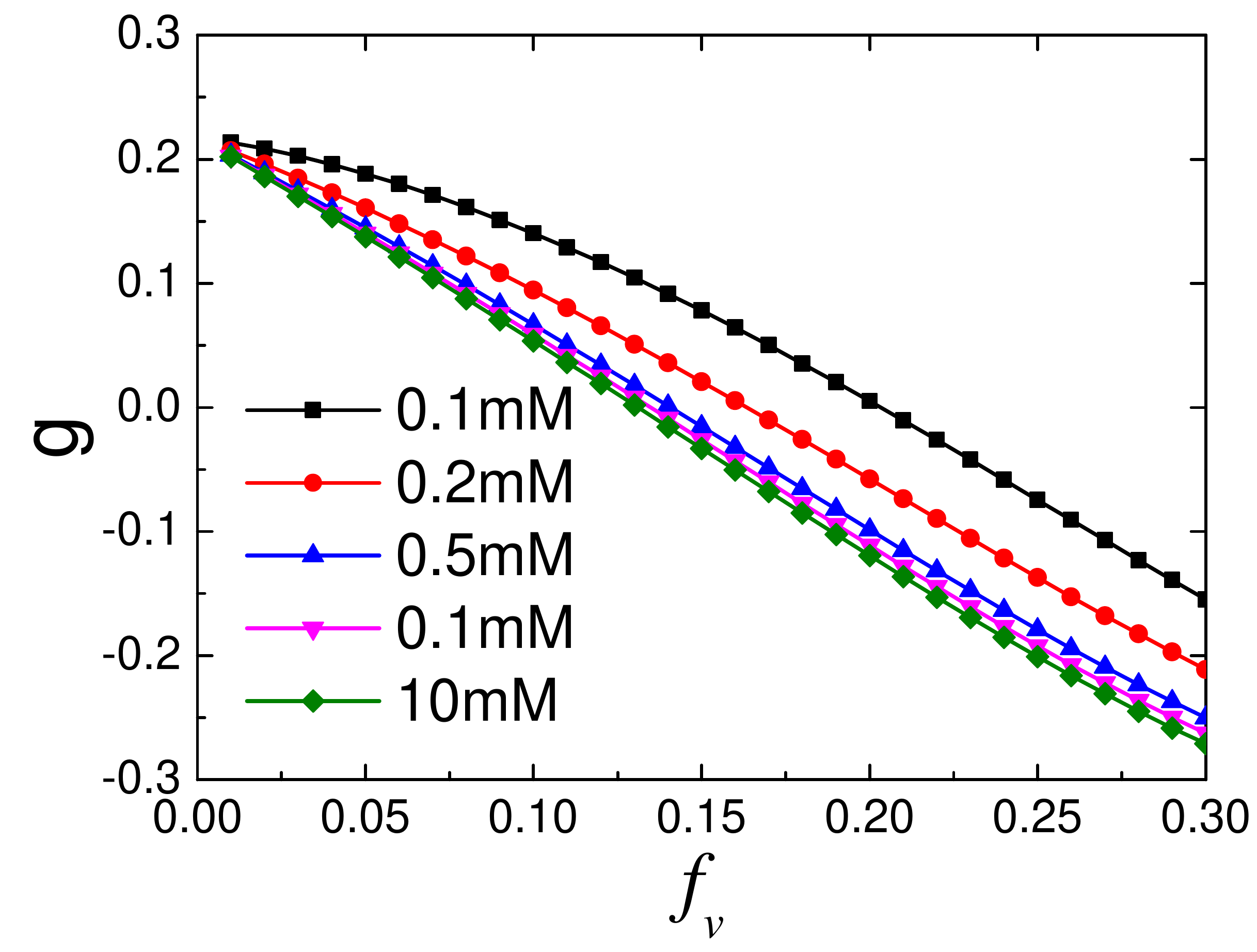}
	}
    \caption{(a) The scattering coefficient $\kappa_s$ and (b) asymmetry factor $g$ as a function of particle volume fraction for PS colloidal suspensions with different salt ion concentrations.}\label{PSZ200}
\end{figure}
\begin{figure}[htbp]
	\centering
	\includegraphics[width=0.8\linewidth]{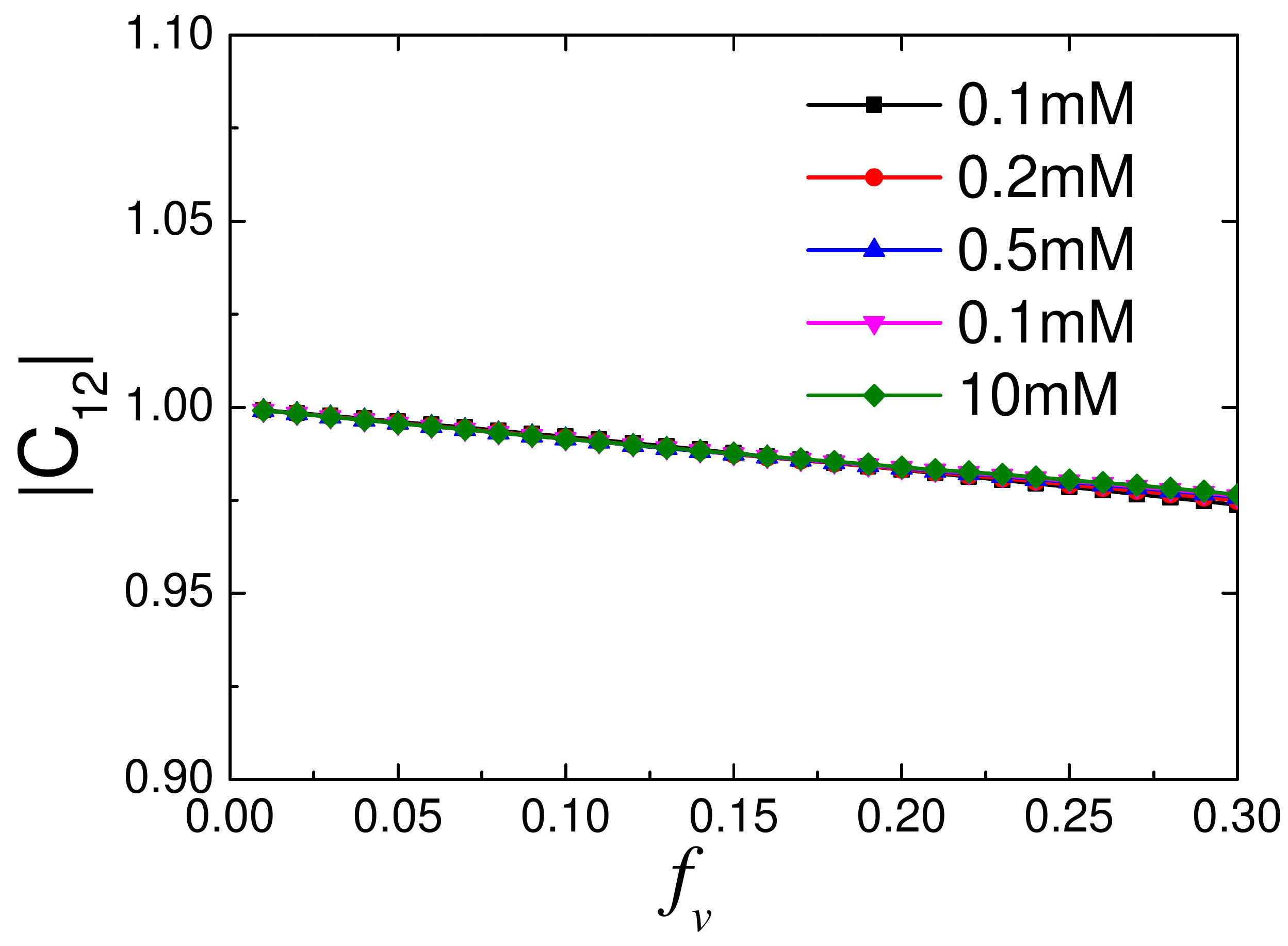}
	\caption{The effective exciting field amplitudes $|C_{12}|$ for electric dipole as a function of particle volume fraction for PS colloidal suspensions with different salt ion concentrations.}\label{CPSZ200}
	\end{figure}

The second system we investigate is the PS colloidal suspensions. The trend of variation of the scattering coefficient and asymmetry factor with particle volume fraction and salt concentration, shown in Fig.\ref{PSZ200}, is similar to the case of $\mathrm{TiO_2}$ suspensions. The reduction of scattering coefficient due to increasing salt concentration is rather small, while the asymmetry factor is substantially lowered. This fact indicates that a lower salt concentration leads to an decrease in transport scattering coefficient $\kappa_\text{tr}=\kappa_\text{tr}(1-g)$, or a growth of transport mean free path $l_\text{tr}=1/\kappa_\text{tr}$. This consequence provides a way to reduce scattering loss in colloidal suspension in the diffusive light transport regime.

However, it is found that for PS colloidal suspensions, the effective field amplitude for electric dipole is almost unchanged over different volume fractions and salt concentrations according to Fig.\ref{CPSZ200}, keeping in the range from 0.97 to 1. Note here in the PS particles only electric dipoles are excited and thus we only calculate $C_{12}$. This is because the refractive index of PS is so low that the screening effect is negligible for the moderate particle volume fraction investigated in our study. Therefore, the dependent scattering mechanism here is only the far-field interference effect described by the structure factor. As a consequence, this is also why for many researches focusing the dependent scattering mechanism on PS colloidal suspensions,  an ITA model is already adequate to explain the experimental results well \cite{fradenPRL1990,rojasochoaPRL2004}. However, for colloidal particles with higher refractive index, this correction from the effective field amplitude should be appropriately taken into consideration.

\section{Conclusions}
To summarize, we study the optical scattering properties of hard-sphere Yukawa (HSY) random media composed of discrete dual-dipolar scatterers, in order to  comprehensively elucidate the role of structural correlations in dependent scattering mechanism. We consider highly like-charged colloidal suspensions as a typical HSY system, for which changing additional salt concentration can flexibly modify the pair interaction between the charged particles and therefore the structural correlations. This strategy can tune the optical scattering properties, including scattering coefficient and asymmetry factor significantly, especially for high refractive index particles like $\mathrm{TiO_2}$. For colloidal $\mathrm{TiO_2}$ suspensions, we demonstrate that the modification of electric and magnetic dipole excitations induced by the structural correlations can substantially affect the optical scattering properties, in addition to the far-field interference effect described by the structure factor. However, this modification is only slightly altered by different salt concentrations and is mainly due to the packing-density-dependent screening effect. On the other hand, for low refractive index colloidal PS suspensions, the dependent scattering mechanism mainly consists of the far-field interference effect, while the effective exciting field amplitude for electric dipole almost remains unchanged under different structural correlations. The present study has profound implications for understanding and exploiting the structural correlations to harness the dependent scattering mechanism and thus optical scattering properties of random media.
\begin{acknowledgments}
We acknowledge the financial support from the National Natural Science Foundation of China (Nos.51636004 and 51476097), Shanghai Key Fundamental Research Grant (16JC1403200) and the Foundation for Innovative Research Groups of the National Natural Science Foundation of China (No.51521004). We also thank Dr. Ming Xiao for fruitful discussions concerning the DLVO theory.
\end{acknowledgments}
\appendix
\section{The MSA solution}\label{MSA}
Here we define the following non-dimensional parameters, $T^*=k_BT/U_0$, which denotes an effective thermodynamic temperature, and $z=\kappa_Dd$, which indicates a normalized inverse screening length. Letting $u=2qa$ and $\Delta=1-f_v$, we list the solutions of $Z_1(u)$ and $Z_2(u)$ under MSA as follows \cite{bresselJSQRT2013}:
\begin{equation}
Z_1(u)=\frac{1}{u^2}[(A-B)+(A+B)\cos u-\frac{2A}{u}\sin u]+zR(u),
\end{equation}
\begin{equation}
Z_2(u)=\frac{1}{u^2}[(u^2Ce^{-u}+2A)\frac{\cos u-1}{u}-Bu+(A+B)\sin u]+uR(u),
\end{equation}
\begin{equation}
R(u)=\frac{u}{u^2+z^2}[\frac{C+D}{z}-Ce^{-u}(\frac{\cos u}{z}+\frac{\sin u}{u})],
\end{equation}
\begin{equation}
A=\frac{2\Gamma}{z\Delta}[\Gamma+z+\frac{1+2f_v}{\Delta}]+\frac{1+2f_v}{\Delta^2},
\end{equation}
\begin{equation}
B=\frac{1}{\Delta}[1-4\frac{\Gamma}{z^2}(\frac{1+2f_v}{\Delta}+\frac{z}{2}+\Gamma)],
\end{equation}
\begin{equation}
D=-\frac{\Gamma}{3f_v}(\Phi_1\Gamma+\Phi_0)e^z,
\end{equation}
\begin{equation}
C=-D+\frac{\Gamma}{3f_v}[\frac{\Gamma}{z}(1-\alpha_1)-\frac{\alpha_0}{z}],
\end{equation}
\begin{equation}
\Phi_1=\phi_0(z)-12\frac{f_v}{\Delta}\psi_1(z),
\end{equation}
\begin{equation}
\Phi_0=1+\frac{3f_v}{\Delta}\phi_0(z)-12\frac{f_v}{\Delta}\psi_1(z)(1+\frac{z}{2}+\frac{3\phi}{\Delta}),
\end{equation}
\begin{equation}
\alpha_0=\alpha_1(\frac{z}{2}+\frac{1+2f_v}{\Delta})-\frac{3f_v}{\Delta},
\end{equation}
\begin{equation}
\alpha_1=\frac{6f_v}{\Delta z^2}(z+2),
\end{equation}
\begin{equation}
\phi_0(z)=\frac{1}{z}(1-e^{-z}),
\end{equation}
\begin{equation}
\Psi_1(z)=\frac{1}{z^3}[1-\frac{z}{2}-(1+\frac{z}{2}e^{-z})].
\end{equation}
Here $\Gamma$ is the root of the following equation:
\begin{equation}
\Gamma^2+z\Gamma=\frac{6f_v}{T^*(\Phi_1\Gamma+\Phi_0)^2}.
\end{equation}
\bibliography{asym_factor_yukawa}

\end{document}